\documentclass[preprint,showpacs,preprintnumbers,amsmath,amssymb]{revtex4}

\usepackage{graphicx}
\usepackage{dcolumn}
\usepackage{bm}
\usepackage{hyperref}
\hypersetup{colorlinks=false}

\begin{document}

\newcommand{\RR}[1]{[#1]}
\newcommand{\intsum}{\sum \kern -15pt \int}
\newfont{\Yfont}{cmti10 scaled 2074}
\newcommand{\Y}{\hbox{{\Yfont y}\phantom.}}
\def\O{{\cal O}}
\newcommand{\bra}[1]{\left< #1 \right| }
\newcommand{\braa}[1]{\left. \left< #1 \right| \right| }
\def\Bra#1#2{{\mbox{\vphantom{$\left< #2 \right|$}}}_{#1}
\kern -2.5pt \left< #2 \right| }
\def\Braa#1#2{{\mbox{\vphantom{$\left< #2 \right|$}}}_{#1}
\kern -2.5pt \left. \left< #2 \right| \right| }
\newcommand{\ket}[1]{\left| #1 \right> }
\newcommand{\kett}[1]{\left| \left| #1 \right> \right.}
\newcommand{\scal}[2]{\left< #1 \left| \mbox{\vphantom{$\left< #1 #2 \right|$}}
\right. #2 \right> }
\def\Scal#1#2#3{{\mbox{\vphantom{$\left<#2#3\right|$}}}_{#1}
{\left< #2 \left| \mbox{\vphantom{$\left<#2#3\right|$}} \right. #3
\right> }}


\title{Three-Dimensional Approach for Construction of Low-Momentum
Effective Interaction from Realistic Potentials}

\author{S. Bayegan}%
\email{bayegan@khayam.ut.ac.ir}
\author{M. Harzchi}
\email{mehdi$_$ harzchi@khayam.ut.ac.ir}
\author{M. A. Shalchi}
\email{shalchi@khayam.ut.ac.ir}

\affiliation{ Department of Physics, University of Tehran, P.O.Box 14395-547, Tehran, Iran}%

\date{\today}

\begin{abstract}
The low-momentum effective interaction $V_{low\,k}$ has been
formulated in the three-dimensional momentum-helicity representation
as a function of the magnitude of momentum vectors and the angle
between them. As an application, AV18 potential has been used in the
model space of Lee-Suzuki method and it has been shown that the
low-momentum effective interaction, $V_{low\,k}$ reproduces the same
two-body observables obtained by the bare potential $V_{NN}$.
\end{abstract}

\maketitle

\section{Introduction} \label{sec:introduction}
Several methods have been developed to derive the energy independent
low-momentum effective interaction such as the renormalization group
(RG) and the model space techniques. These approaches are mainly
based on a partial wave (PW) decomposition and the details have been
given in references \cite{Epelbaum-PLB439}-\cite{Bogner-PLB576}.
Recently Bogner \emph{et al}. have developed a low-momentum
effective interaction which describes the two-nucleon system at low
energy successfully. This effective interaction is independent of
the potential models as the cutoff is lowered to
$\Lambda=2.1\,\,fm^{-1}$ \cite{Bogner-PR386,Bogner-PLB576}.

During the past years, the three-dimensional (3D) approach has been
developed for few-body bound and scattering problems
\cite{Elster-FBS24}-\cite{Fachruddin-PRC68}. In this approach
momentum helicity basis states have been used for the representation
of the nuclear forces. The motivation for developing this approach
is to introduce a direct solution of the integral equations avoiding
the very involved angular momentum algebra occurring for the
permutations, transformations and especially for the three-body
forces. Conceptually the 3D formalism considers all partial wave
channels automatically.

Recently we have developed a 3D formalism for construction of
low-momentum effective interaction neglecting the spin and isospin
degrees of freedom \cite{Bayegan-p}. Considering the spin and
isospin is a major additional task, which we intend to present in
this article. Our aim is to extend the low-momentum effective
interaction directly in a spin-isospin dependent 3D approach and to
formulate the low-momentum effective interaction in the
momentum-helicity representation.

This article is structured in the following way: In section
\ref{sec:Lee-Suzuki method} the model space of Lee-Suzuki method has
been used to derive the energy independent model space effective
interaction in the 3D momentum-helicity representation. In section
\ref{sec:coordinate systems} the reduced forms of the equations have
been displayed by choosing suitable coordinate systems. Section
\ref{sec:Numerical results} describes the numerical calculations of
the low-momentum effective interaction $V_{low\,k}$ in the model
space Lee-Suzuki method by using the AV18 potential. Finally a
summary and an outlook have been provided in section
\ref{sec:summary}.

\section{Lee-Suzuki method in the 3D momentum-helicity representation}\label{sec:Lee-Suzuki method}
The Lee-Suzuki method has been applied to the free space
nucleon-nucleon problem in the 3D momentum-helicity representation
and the low-momentum effective interaction $V_{low\,k}$ has been
obtained as a function of the momentum vectors. In the model space
methods the projection operators onto the physically important
low-energy model space, the $P$ space, and the high-energy
complement, the $Q$ space, have been introduced on momentum-helicity
basis state $|\textbf{k};\textbf{\^{k}}S\lambda;t\rangle^{\pi a}$
as:
\begin{eqnarray}
\nonumber P &=& \sum_{S\lambda\pi t}\int
d\textbf{k}\,|\textbf{k};\textbf{\^{k}}S\lambda;t\rangle^{\pi
a}\,\frac{1}{4}\,\ ^{\pi
a}\langle\textbf{k};\textbf{\^{k}}S\lambda;t|,\,\,\,\,|\textbf{k}|\leq
\Lambda, \\ Q &=& \sum_{S\lambda\pi t}\int
d\textbf{k}\,|\textbf{k};\textbf{\^{k}}S\lambda;t\rangle^{\pi
a}\,\frac{1}{4}\,\ ^{\pi
a}\langle\textbf{k};\textbf{\^{k}}S\lambda;t|,\,\,\,\,|\textbf{k}|>\Lambda,
\end{eqnarray}
where $\Lambda$ is a momentum cutoff which divides the Hilbert space
into the low and high momentum states. The antisymmetrized
momentum-helicity basis state which is parity eigenstate is given by
\cite{Fachruddin-PRC62}:
\begin{eqnarray} \label{momentum-helicity basis}
|\textbf{k};\textbf{\^{k}}S\lambda;t\rangle^{\pi a}
&=&\frac{1}{\sqrt{2}}(1-P_{12})|\textbf{k};\textbf{\^{k}}S\lambda\rangle_{\pi}\,|t\rangle\nonumber
\\&=&\frac{1}{\sqrt{2}}(1-\eta_{\pi}(-)^{S+t})|\textbf{k};\textbf{\^{k}}S\lambda\rangle_{\pi}\,|t\rangle,
\end{eqnarray}
Here $S$ is the total spin, $\lambda$ is the spin projection along
relative momentum of two nucleons, $t$ is the total isospin and
$|t\rangle\equiv|tm_{t}\rangle$ is the total isospin state of the
two nucleons. $m_{t}$ is the isospin projection along its
quantization axis which reveals the total electric charge of system.
For simplicity $m_{t}$ is suppressed since electric charge is
conserved. In Eq. (\ref{momentum-helicity basis}) $P_{12}$ is the
permutation operator which exchanges the two nucleons labels in all
spaces i.e. momentum, spin and isospin spaces, and
$|\textbf{k};\textbf{\^{k}}S\lambda\rangle_{\pi}$ is parity
eigenstate  which is given by:
\begin{eqnarray}
|\textbf{k};\textbf{\^{k}}S\lambda\rangle_{\pi}=\frac{1}{\sqrt{2}}(1+\eta_{\pi}P_{\pi})|\textbf{k};
\textbf{\^{k}}S\lambda\rangle,
\end{eqnarray}
where $P_{\pi}$ is parity operator, $\eta_{\pi}=\pm1$ are the parity
eigenvalues and $|\textbf{k};\textbf{\^{k}}S\lambda\rangle$ is
momentum-helicity state. The normalization of the momentum-helicity
basis state is given by \cite{Fachruddin-PRC62}:
\begin{eqnarray}
\,^{\pi'a}\langle\textbf{k}';\textbf{\^{k}}'S'\lambda';t'|\textbf{k};\textbf{\^{k}}S\lambda;\,t\rangle^{\pi
a}
&=&(1-\eta_{\pi}(-)^{S+t})\,\delta_{t't}\delta_{\eta_{\pi'}\eta_{\pi}}\delta_{S'S}\nonumber
\\&\times&\{\delta(\textbf{k}'-\textbf{k})\delta_{\lambda'\lambda}+\eta_{\pi}(-)^{S}\delta(\textbf{k}'+
\textbf{k})\delta_{\lambda'-\lambda}\},\nonumber\\
\end{eqnarray}
and the completeness relation of this state is defined by:
\begin{eqnarray}
\sum_{S\lambda\pi t}\int
d\textbf{k}\,|\textbf{k};\textbf{\^{k}}S\lambda;t\rangle^{\pi
a}\,\frac{1}{4}\ ^{\pi
a}\langle\textbf{k};\textbf{\^{k}}S\lambda;t|=1.
\end{eqnarray}
It is clear that the projection operators $P$ and $Q$ satisfy the
following relations:
\begin{eqnarray}
P+Q &=& 1, \nonumber \\  PQ&=&QP=0, \nonumber \\  P^{2}&=&P,
\nonumber \\  Q^{2}&=&Q,
\end{eqnarray}
and they act on the full-space two-body problem states as:
\begin{eqnarray}
\nonumber P|\psi_{\textbf{k};\textbf{\^{k}}S\lambda;t}\rangle^{\pi
a} &=& |\Psi_{\textbf{k};\textbf{\^{k}}S\lambda;t}\rangle^{\pi a},
\\
Q|\psi_{\textbf{k};\textbf{\^{k}}S\lambda;t}\rangle^{\pi a} &=&
\omega|\Psi_{\textbf{k};\textbf{\^{k}}S\lambda;t}\rangle^{\pi a},
\end{eqnarray}
where $|\psi_{\textbf{k};\textbf{\^{k}}S\lambda;t}\rangle^{\pi a}$
and $|\Psi_{\textbf{k};\textbf{\^{k}}S\lambda;t}\rangle^{\pi a}$
denote the states of the full and model spaces respectively and
$\omega$ is an operator which transforms the states of the $P$-space
to the $Q$-space. The non-hermitian low-momentum effective potential
in the model space that reproduces the model space component of the
wave function from the full-space wave function is given by
\cite{Suzuki-PTP68}-\cite{Suzuki-PTP70}:
\begin{eqnarray}
V_{low\,k}=PV_{NN}(P+Q\omega P),
\end{eqnarray}
where $V_{NN}$ denotes the bare two-body interaction. By using the
integral form of the projection operators $P$ and $Q$, the
low-momentum effective interaction $V_{low\,k}$ can be written in
the 3D representation on momentum-helicity basis state as:
\begin{eqnarray} \label{eq.vlowk1}
^{\pi
a}\langle\textbf{p}';\textbf{\^{p}}'S\lambda';t|V_{low\,k}|\textbf{p};\textbf{\^{p}}S\lambda;t\rangle^{\pi
a}&=&\,^{\pi a}\langle\textbf{p}';\textbf{\^{p}}'S\lambda';t|V_{NN}|\textbf{p};\textbf{\^{p}}S\lambda;t
\rangle^{\pi a} \nonumber\\&+&\frac{1}{4}\sum_{\lambda''}\int_{\Lambda}^{\infty}
dq\,q^{2}\int d\hat{\textbf{q}} \,\,^{\pi
a}\langle\textbf{p}';\textbf{\^{p}}'S\lambda';t|V_{NN}|
\textbf{q};\textbf{\^{q}}S\lambda'';t\rangle^{\pi
a}\nonumber\\&\times& \,^{\pi
a}\langle\textbf{q};\textbf{\^{q}}S\lambda'';t|\omega|\textbf{p};\textbf{\^{p}}S\lambda;t
\rangle^{\pi a},
\end{eqnarray}
where we use the property that $V_{NN}$ conserves parity, spin and
isospin. By defining matrix elements of each arbitrary operator $A$
as:
\begin{eqnarray}
A^{\pi St;\lambda'\lambda}(\textbf{p}',\textbf{p})\equiv\,^{\pi
a}\langle\textbf{p}';\textbf{\^{p}}'S\lambda';t|A|\textbf{p};\textbf{\^{p}}S\lambda;t\rangle^{\pi
a}.
\end{eqnarray}
Eq. (\ref{eq.vlowk1}) can be rewritten as:
\begin{eqnarray} \label{eq.vlowk}
V^{\pi St,\lambda'\lambda}_{low\,k}(\textbf{p}',\textbf{p})= V^{\pi
St,\lambda'\lambda}_{NN}(\textbf{p}',\textbf{p})+
\frac{1}{4}\sum_{\lambda''}\int_{\Lambda}^{\infty} dq\,q^{2}\int
d\hat{\textbf{q}}\,\,V^{\pi
St,\lambda'\lambda''}_{NN}(\textbf{p}',\textbf{q})\,\,\omega^{\pi
St,\lambda''\lambda}(\textbf{q},\textbf{p}),
\end{eqnarray}
where \textbf{p} and \textbf{q} are the momentum vectors in the $P$
and $Q$ spaces respectively. As we know, the total spin states of
two nucleons are singlet $(S=0)$ and triplet $(S=1)$ states. For the
singlet case, Eq. (\ref{eq.vlowk}), there is only one equation which
reads as:
\begin{eqnarray}
V^{\pi 0t,00}_{low\,k}(\textbf{p}',\textbf{p})&=& V^{\pi
0t,00}_{NN}(\textbf{p}',\textbf{p})+
\frac{1}{4}\int_{\Lambda}^{\infty} dq\,q^{2}\int
d\hat{\textbf{q}}\,\,V^{\pi
0t,00}_{NN}(\textbf{p}',\textbf{q})\,\,\omega^{\pi
0t,00}(\textbf{q},\textbf{p}),
\end{eqnarray}
however, for the triplet case there are three equations for each
initial helicity $\lambda=-1,0,1$ as:
\begin{eqnarray} \label{eq.vlowk-triplet}
V^{\pi 1t,\lambda'\lambda}_{low\,k}(\textbf{p}',\textbf{p})&=&
V^{\pi 1t,\lambda'\lambda}_{NN}(\textbf{p}',\textbf{p})+\frac{1}{4}\int_{\Lambda}^{\infty}dq\,q^{2}\int
d\hat{\textbf{q}}\,\{V^{\pi
1t,\lambda'1}_{NN}(\textbf{p}',\textbf{q})\,\omega^{\pi
1t,1\lambda}(\textbf{q},\textbf{p})\nonumber\\ &+&V^{\pi
1t,\lambda'0}_{NN}(\textbf{p}',\textbf{q}) \,\omega^{\pi
1t,0\lambda}(\textbf{q},\textbf{p})+V^{\pi
1t,\lambda'-1}_{NN}(\textbf{p}',\textbf{q}) \,\omega^{\pi
1t,-1\lambda}(\textbf{q},\textbf{p})\}.
\end{eqnarray}
Symmetry relations for the potential matrix elements in the
momentum-helicity representation are denoted by
\cite{Fachruddin-PRC62}:
\begin{eqnarray}
V^{\pi \nonumber
St,-\lambda'\lambda}_{NN}(\textbf{p}',\textbf{p})&=&\eta_{\pi}(-)^{S}V^{\pi
St,\lambda'\lambda}_{NN}(-\textbf{p}',\textbf{p}), \nonumber\\V^{\pi
St,\lambda'-\lambda}_{NN}(\textbf{p}',\textbf{p})&=&\eta_{\pi}(-)^{S}V^{\pi
St,\lambda'\lambda}_{NN}(\textbf{p}',-\textbf{p}),\nonumber\\V^{\pi
St,-\lambda'-\lambda}_{NN}(\textbf{p}',\textbf{p})&=&V^{\pi
St,\lambda'\lambda}_{NN}(-\textbf{p}',-\textbf{p}).
\end{eqnarray}
A corresponding symmetry relations are also valid for
$\omega$-matrix elements. By applying this symmetry relations for
matrix elements of $V_{NN}$ and $\omega$, Eq.
(\ref{eq.vlowk-triplet}) can be reduced to:
\begin{eqnarray} \label{eq.Vlowk-triplet}
V^{\pi 1t,\lambda'\lambda}_{low\,k}(\textbf{p}',\textbf{p})&=&
V^{\pi 1t,\lambda'\lambda}_{NN}(\textbf{p}',\textbf{p})+\frac{1}{4}\int_{\Lambda}^{\infty}
dq\,q^{2}\int d\hat{\textbf{q}}\,\,\{2\,V^{\pi
1t,\lambda'1}_{NN}(\textbf{p}',\textbf{q})\,\omega^{\pi
1t,1\lambda}(\textbf{q},\textbf{p})\nonumber \\&+& V^{\pi
1t,\lambda'0}_{NN}(\textbf{p}',\textbf{q})\,\omega^{\pi
1t,0\lambda}(\textbf{q},\textbf{p})\}.
\end{eqnarray}
Hence for the triplet case one needs only two equations for
$\lambda'=1,0$ for each $\lambda$. To calculate the low-momentum
effective potential $V_{low\,k}$ we need to determine
$\omega$-matrix elements. The key aspect of the Lee-Suzuki method is
the determination of the $\omega$ operator defined by the following
equation \cite{Jennings-EL72}:
\begin{eqnarray}
Q|\psi_{\textbf{k};\textbf{\^{k}}S\lambda;t}\rangle^{\pi
a}=Q\,\omega
P|\psi_{\textbf{k};\textbf{\^{k}}S\lambda;t}\rangle^{\pi a}.
\end{eqnarray}
Applying $\,^{\pi a}\langle\textbf{q};\textbf{\^{q}}S\lambda';t|$ to
the left hand side and using the integral form of projection
operators $P$ and $Q$, this equation can be rewritten as:
\begin{eqnarray} \label{eq.omega}
^{\pi
a}\langle\textbf{q};\textbf{\^{q}}S\lambda';t|\Psi_{\textbf{k};\textbf{\^{k}}S\lambda;t}\rangle^{\pi
a}&=&\frac{1}{4}\sum_{\lambda''}\int_{0}^{\Lambda} dp\,p^{2}\int
d\hat{\textbf{p}}\, ^{\pi a}\langle\textbf{q};
\textbf{\^{q}}S\lambda';t|\omega|
\textbf{p};\textbf{\^{p}}S\lambda'';t\rangle^{\pi a} \nonumber\\
&\times&\,^{\pi
a}\langle\textbf{p};\textbf{\^{p}}S\lambda'';t|\Psi_{\textbf{k};\textbf{\^{k}}S
\lambda;t}\rangle^{\pi a}.
\end{eqnarray}
We introduce the completeness relation for the scattering wave
function in the model space on momentum-helicity basis state:
\begin{eqnarray} \label{eq.completness}
&&\frac{1}{4}\sum_{\lambda''}\int_{0}^{\Lambda}dk\,k^{2}\int
d\,\hat{\textbf{k}}\,\Psi^{\pi
St,\lambda'\lambda''}_{\textbf{k}}}(\textbf{p}')
{\,\tilde{\Psi}_{\textbf{k}}^{\pi St,\lambda''\lambda}(\textbf{p})
\nonumber\\&&=(1-\eta_{\pi}(-)^{S+t})\{\delta(\textbf{p}'-\textbf{p})\delta_{\lambda'\lambda}+
\eta_{\pi}(-)^{S}\delta(\textbf{p}'+\textbf{p})\delta_{\lambda'-\lambda}\},
\end{eqnarray}
where we use the following notations for the matrix elements of the
wave functions:
\begin{eqnarray}
\Psi^{\pi
St,\lambda'\lambda}_{\textbf{k}}(\textbf{p})&\equiv&\,^{\pi
a}\langle\textbf{p};\textbf{\^{p}}S\lambda';t|\Psi_{\textbf{k};\textbf{\^{k}}S\lambda;t}\rangle^{\pi
a},\nonumber \\
\tilde{\Psi}_{\textbf{k}}^{\pi
St,\lambda\lambda'}(\textbf{p})&\equiv&\,^{\pi
a}\langle\tilde{\Psi}_{\textbf{k};\textbf{\^{k}}S\lambda;t}|\textbf{p};\textbf{\^{p}}S\lambda';t\rangle^{\pi
a}.
\end{eqnarray}
After implementing Eq. (\ref{eq.completness}), Eq. (\ref{eq.omega})
can be written as:
\begin{eqnarray} \label{eq.omega0}
&&\omega^{\pi
St,\lambda'\lambda}(\textbf{q},\textbf{p})=\frac{1}{4}\sum_{\lambda''}\int_{0}^{\Lambda}
dk\,k^{2}\int d\hat{\textbf{k}}\,\,\Psi_{\textbf{k}}^{\pi
St,\lambda'\lambda''}(\textbf{q})\,\tilde{\Psi}_{\textbf{k}}^{\pi
St,\lambda''\lambda}(\textbf{p}) ,
\end{eqnarray}
where $\Psi^{\pi St,\lambda'\lambda}_{\textbf{k}}(\textbf{p})$ and
$\Psi^{\pi St,\lambda'\lambda}_{\textbf{k}}(\textbf{q})$ are the
wave function components of the $P$ and $Q$ spaces of the
full-space, respectively. These can be written in the form of the
half-on-shell (HOS) two-body $T$-matrix as:
\begin{eqnarray}
\Psi^{\pi St,\lambda'\lambda}_{\textbf{k}}(\textbf{p})=
(1-\eta_{\pi}(-)^{S+t})\{\delta(\textbf{p}-
\textbf{k})\delta_{\lambda'\lambda}+\eta_{\pi}(-)^{S}
\delta(\textbf{p}+\textbf{k})
\delta_{\lambda'-\lambda}+\frac{1}{2}\frac{T^{\pi
St,\lambda'\lambda}(\textbf{p},
\textbf{k},k^{2})}{k^{2}-p^{2}+i\varepsilon}\},\nonumber \\
\end{eqnarray}
\begin{eqnarray}
\Psi^{\pi
St,\lambda'\lambda}_{\textbf{k}}(\textbf{q})&=&\frac{T^{\pi
St,\lambda'\lambda}(\textbf{q},\textbf{k},k^{2})}{k^{2}-q^{2}}.
\end{eqnarray}
The HOS two-body $T$-matrix can be obtained from the
Lippmann-Schwinger (LS) equation in the 3D momentum-helicity
representation, which is given by $(\hbar=m_{N}=1)$
\cite{Fachruddin-PRC62}:
\begin{eqnarray}
T^{\pi St,\lambda'\lambda}(\textbf{k}',\textbf{k},k^{2})&=&V^{\pi
St,\lambda'\lambda}_{NN}(\textbf{k}',\textbf{k})
+\frac{1}{4}\sum_{\lambda''}\int\emph{d}\textbf{k}'' \frac{V^{\pi
St,\lambda'\lambda''}_{NN}(\textbf{k}',\textbf{k}'')\,T^{\pi
St,\lambda''\lambda}(\textbf{k}'',\textbf{k},k^{2})}{k^{2}-k''^{2}+i\varepsilon},\nonumber \\
\end{eqnarray}
The equation for the singlet case is:
\begin{eqnarray}
T^{\pi 0t,00}(\textbf{k}',\textbf{k},k^{2})&=&V^{\pi
0t,00}_{NN}(\textbf{k}',\textbf{k})+\frac{1}{4}\int\emph{d}\textbf{k}''
\frac{V^{\pi 0t,00}_{NN}(\textbf{k}',\textbf{k}'')\,T^{\pi
0t,00}(\textbf{k}'',\textbf{k},k^{2})}{k^{2}-k''^{2}+i\varepsilon},
\end{eqnarray}
and for the triplet case we have two coupled equations for
$\lambda'=1,0$ for each $\lambda$ as:
\begin{eqnarray}
T^{\pi 1t,\lambda'\lambda}(\textbf{k}',\textbf{k},k^{2})&=&V^{\pi
1t,\lambda'\lambda}_{NN}(\textbf{k}',\textbf{k})
+\frac{1}{2}\int\emph{d}\textbf{k}'' \frac{V^{\pi
1t,\lambda'1}_{NN}(\textbf{k}',\textbf{k}'')\,T^{\pi
1t,1\lambda}(\textbf{k}'',\textbf{k},k^{2})}{k^{2}-k''^{2}+i\varepsilon}\nonumber
\\&+&\frac{1}{4}\int\emph{d}\textbf{k}'' \frac{V^{\pi
1t,\lambda'0}_{NN}(\textbf{k}',\textbf{k}'')\,T^{\pi
1t,0\lambda}(\textbf{k}'',\textbf{k},k^{2})}{k^{2}-k''^{2}+i\varepsilon}.
\end{eqnarray}
Eqs. (\ref{eq.completness}) and (\ref{eq.omega0}) for the singlet
case can be written as:
\begin{eqnarray}
\frac{1}{4}\int_{0}^{\Lambda}dk\,k^{2}\int d\,\hat{\textbf{k}}\,\,
\Psi^{\pi 0t,00}_{\textbf{k}}}(\textbf{p}')
{\tilde{\Psi}_{\textbf{k}}^{\pi 0t,00}(\textbf{p})
=(1-\eta_{\pi}(-)^{t})\{\delta(\textbf{p}'-\textbf{p})+\eta_{\pi}\delta(\textbf{p}'+\textbf{p})\},
\end{eqnarray}
\begin{eqnarray}
&&\omega^{\pi
0t,00}(\textbf{q},\textbf{p})=\frac{1}{4}\int_{0}^{\Lambda}dk\,k^{2}\int
d\hat{\textbf{k}} \,\,\Psi_{\textbf{k}}^{\pi
0t,00}(\textbf{q})\,\tilde{\Psi}_{\textbf{k}}^{\pi
0t,00}(\textbf{p}).
\end{eqnarray}
It is clear that the same symmetry relations as we mentioned for
potential matrix elements are valid for matrix elements of $\Psi$
and $\tilde{\Psi}$. Thus by applying the symmetry relations for
triplet case, we write two equations for $\lambda'=1,0$ for each
$\lambda$ respectively:
\begin{eqnarray}
&&\frac{1}{4}\int_{0}^{\Lambda}dk\,k^{2}\int
d\,\hat{\textbf{k}}\,\,\{2\,\Psi_{\textbf{k}}^{\pi
1t,\lambda'1}(\textbf{p}')\,\tilde{\Psi}^{\pi
1t,1\lambda}_{\textbf{k}}}(\textbf{p})\,+ { \,\Psi_{\textbf{k}}^{\pi
1t,\lambda'0\overline{}}(\textbf{p}')\tilde{\Psi}^{\pi
1t,0\lambda}_{\textbf{k}}}(\textbf{p}) {
\,\}\nonumber\\&&=(1+\eta_{\pi}(-)^{t})\{\delta(\textbf{p}'-\textbf{p})\delta_{\lambda'\lambda}-
\eta_{\pi}\delta(\textbf{p}'+\textbf{p})\delta_{\lambda'-\lambda}\},
\end{eqnarray}
\begin{eqnarray}
\omega^{\pi
1t,\lambda'\lambda}(\textbf{q},\textbf{p})&=&\frac{1}{4}\int_{0}^{\Lambda}
dk\,k^{2}\int d\hat{\textbf{k}}\,\{2\,\Psi_{\textbf{k}}^{\pi
1t,\lambda'1}(\textbf{q})\, \tilde{\Psi}_{\textbf{k}}^{\pi
1t,1\lambda}(\textbf{p})+\,\Psi_{\textbf{k}}^{\pi 1t,\lambda'0}
(\textbf{q})\,\tilde{\Psi}_{\textbf{k}}^{\pi
1t,0\lambda}(\textbf{p})\}.\nonumber \\
\end{eqnarray}
\section{Choosing suitable coordinate systems for numerical calculations}\label{sec:coordinate systems}
In this section suitable coordinate systems are chosen in order to
write the Eqs. (\ref{eq.vlowk}), (\ref{eq.completness}) and
(\ref{eq.omega0}) in reduced forms for numerical calculations. The
azimuthal behavior of potential and two-body $T$-matrix elements for
special case where the vector $k$ is along $z$ axis are given by
\cite{Fachruddin-PRC62}:
\begin{eqnarray}
T^{\pi
St,\lambda'\lambda}(\textbf{k}',k\hat{\textbf{z}})&=&e^{i\lambda\varphi'}T^{\pi
St,\lambda'\lambda}(k',k,x'),\nonumber \\V_{NN}^{\pi
St,\lambda'\lambda}(\textbf{k}',k\hat{\textbf{z}})&=&
e^{i\lambda\varphi'}V_{NN}^{\pi St,\lambda'\lambda}(k',k,x'),
\end{eqnarray}
where $x'=\hat{\textbf{k}}'\cdot\hat{\textbf{k}}$. Inserting these
relations into the LS equation one can obtain
\cite{Fachruddin-PRC62}:
\begin{eqnarray} \label{eq.T-matrix}
&&T^{\pi St,\lambda'\lambda}(k',k,x',k^{2})=V_{NN}^{\pi
St,\lambda'\lambda}(k',k,x')
\nonumber\\&&+\frac{1}{4}\sum_{\lambda''}\int_{0}^{\infty}dk''k''\int_{-1}^{1}dx''\,\,\frac{\nu^{\pi
St,\lambda'\lambda''}_{\lambda}(k',k'',x',x'')\,\,T^{\pi
St,\lambda''\lambda}(k'',k,x'',k^{2})}{k^{2}-k''^{2}+i\varepsilon}.
\end{eqnarray}
where $x''=\hat{\textbf{k}}''\cdot\hat{\textbf{k}}$ and
$\varphi''$-integration is carried out independently by defining:
\begin{eqnarray}
\nu^{\pi
St,\lambda'\lambda''}_{\lambda}(k',k'',x',x'')\equiv\int_{0}^{2\pi}d\varphi''\,\,
e^{-i\lambda(\varphi'-\varphi'')}\,\,V_{NN}^{\pi
St,\lambda'\lambda''}(\textbf{k}',\textbf{k}'').
\end{eqnarray}
The integrand is periodical with respect to $\varphi''$. Thus we can
set $\varphi'=0$ and Eq. (\ref{eq.T-matrix}) for the singlet and the
triplet cases can be written as:
\begin{eqnarray} \label{eq.T-matrix1}
T^{\pi 0t,00}&&(k',k,x',k^{2})=V_{NN}^{\pi 0t,00}(k',k,x')\nonumber\\
&&+\frac{1}{4}\int_{0}^{\infty}dk''k''\int_{-1}^{1}dx''\,\frac{\nu^{\pi
0t,00}_{0}(k',k'',x',x'')T^{\pi
0t,00}(k'',k,x'',k^{2})}{k^{2}-k''^{2}+i\varepsilon},
\end{eqnarray}
\begin{eqnarray} \label{eq.T-matrix2}
T^{\pi 1t,\lambda'\lambda}&&(k',k,x',k^{2})= V_{NN}^{\pi
1t,\lambda'\lambda}(k',k,x')
\nonumber\\&&+\frac{1}{2}\int_{0}^{\infty}dk''k''\int_{-1}^{1}dx''\,
\frac{\nu^{\pi 1t,\lambda'1}_{\lambda}(k',k'',x',x'')T^{\pi
1t,1\lambda}(k'',k,x'',k^{2})}{k^{2}-k''^{2}+i\varepsilon}\nonumber\\&&
+\frac{1}{4}\int_{0}^{\infty}dk''k''\int_{-1}^{1}dx''\,
\frac{\nu^{\pi 1t,\lambda'0}_{\lambda}(k',k'',x',x'')T^{\pi
1t,0\lambda}(k'',k,x'',k^{2}}{k^{2}-k''^{2}+i\varepsilon}).
\end{eqnarray}
As we know the $T$-matrix elements $T^{\pi
St,\lambda'\lambda}(\textbf{k}',\textbf{k},k^{2})$ are not the
solution of the LS equation. The solution of the LS equation would
be $T^{\pi St,\lambda'\lambda}(k',k,x',k^{2})$, which are $T$-matrix
elements in momentum-helicity basis with initial momentum in the
$z$-direction and without its azimuthal dependence. Therefore
$T^{\pi St,\lambda'\lambda}(\textbf{k}',\textbf{k},k^{2})$ can be
connected to the solution of the LS equation as follows
\cite{Fachruddin-PRC68}:
\begin{eqnarray}
T^{\pi
St,\lambda'\lambda}(\textbf{k}',\textbf{k},k^{2})=\frac{\sum_{N=-S}^{S}e^{iN(\varphi'-\varphi)}\,
d^{S}_{N\lambda'}(x')\,d^{S}_{N\lambda}(x)}{\,d^{S}_{\lambda'\lambda}(y)}T^{\pi
St,\lambda'\lambda}(k',k,y,k^{2}),
\end{eqnarray}
where:
\begin{eqnarray}
y&=&\hat{\textbf{k}}'\cdot\hat{\textbf{k}}=x'x+\sqrt{1-x'^{2}}\sqrt{1-x^{2}}\cos(\varphi'-\varphi),
\end{eqnarray}
and $d^{S}_{\lambda'\lambda}(x)$ are rotation matrices \cite{Rose}.
Now we consider the azimuthal behavior of $\Psi$ and $\omega$-matrix
elements:
\begin{eqnarray}
\omega^{\pi
St,\lambda'\lambda}(q\hat{\textbf{z}},\textbf{p})&=&e^{-i\lambda'\varphi}\omega^{\pi
St,\lambda'\lambda}(q,p,x),\nonumber \\\Psi_{\textbf{p}}^{\pi
St,\lambda'\lambda}(q\hat{\textbf{z}})&=&
e^{-i\lambda'\varphi}\Psi_{p}^{\pi St,\lambda'\lambda}(q,x),
\end{eqnarray}
where the vector $\textbf{q}$ is along the $z$ axis and
$x=\hat{\textbf{p}}\cdot\hat{\textbf{q}}$. The Eq. (\ref{eq.omega0})
can be reduced to:
\begin{eqnarray} \label{eq.omega1}
\omega^{\pi
St,\lambda'\lambda}(q,p,x)=\frac{1}{4}\sum_{\lambda''}\int_{0}^{\Lambda}dk\,k^{2}
\int_{-1}^{1}dx''\,\,\Psi^{\pi St,
\lambda'\lambda''}_{k}(q,x'')\,\,\bar{\tilde{\Psi}}^{\pi
St,\lambda''\lambda}_{k,\lambda'}(p,x'',x),
\end{eqnarray}
where $x''=\hat{\textbf{k}}\cdot\hat{\textbf{q}}$ and:
\begin{eqnarray}
\Psi^{\pi St,\lambda'\lambda''}_{k}(q,x'')&=&\frac{T^{\pi
St,\lambda'\lambda''}(q,k,x'',k^{2})}{k^{2}-q^{2}}.
\end{eqnarray}
Integration over $\varphi''$ can be performed independently by
defining:
\begin{eqnarray}
\bar{\tilde{\Psi}}^{\pi
St,\lambda''\lambda}_{k,\lambda'}(p,x'',x)\equiv\int_{0}^{2
\pi}d\varphi''\, e^{-i\lambda'(\varphi''-\varphi)}\tilde{\Psi}^{\pi
St,\lambda''\lambda}_{\textbf{k}}(\textbf{p}).
\end{eqnarray}
Eq. (\ref{eq.omega1}) for the singlet and the triplet cases can be
written as:
\begin{eqnarray} \label{eq.omega2}
\omega^{\pi 0t,00}(q,p,x)=\frac{1}{4}\int_{0}^{\Lambda}dk\,k^{2}
\int_{-1}^{1}dx''\,\,\Psi^{\pi
0t,00}_{k}(q,x'')\,\,\bar{\tilde{\Psi}}^{\pi 0t,00}_{k,0}(p,x'',x),
\end{eqnarray}
\begin{eqnarray} \label{eq.omega3}
\omega^{\pi
1t,\lambda'\lambda}(q,p,x)&=&\frac{1}{4}\int_{0}^{\Lambda}dk\,k^{2}
\int_{-1}^{1}dx''\,\{2\,\Psi^{\pi 1t,
\lambda'1}_{k}(q,x'')\,\,\bar{\tilde{\Psi}}^{\pi
1t,1\lambda}_{k,\lambda'}(p,x'',x)\nonumber \\&+&\Psi^{\pi 1t,
\lambda'0}_{k}(q,x'')\,\,\bar{\tilde{\Psi}}^{\pi
1t,0\lambda}_{k,\lambda'}(p,x'',x)\,\}.
\end{eqnarray}
Multiplying both sides of Eq. (\ref{eq.completness}) by
$e^{-i\lambda'''(\varphi'-\varphi)}$ and integrating over $\varphi'$
yields:
\begin{eqnarray} \label{eq.completness1}
&&\frac{1}{4}\sum_{\lambda''}\int_{0}^{\Lambda}dk\,k^{2}\int_{-1}^{1}
d\,x''\,\bar{\Psi}^{\pi
St,\lambda'\lambda''}_{k,\lambda'''}}(p',x',x'') {
\,\,\bar{\tilde{\Psi}}_{k,\lambda'''}^{\pi
St,\lambda''\lambda}(p,x'',x)
\nonumber\\&&=(1-\eta_{\pi}(-)^{S+t})\,\delta(p'-p)\{\delta(x'-x)
\delta_{\lambda'\lambda}+\eta_{\pi}(-)^{S+\lambda'''}\delta(x'+x)\delta_{\lambda'-\lambda}\},
\end{eqnarray}
where:
\begin{eqnarray}
\bar{\Psi}^{\pi
St,\lambda'\lambda''}_{k,\lambda'''}(p',x',x'')\equiv\int_{0}^{2
\pi}d\varphi'\, e^{-i\lambda'''(\varphi'-\varphi'')}\Psi^{\pi
St,\lambda'\lambda''}_{\textbf{k}}(\textbf{p}'),
\end{eqnarray}
and:
\begin{eqnarray}
\Psi_{\textbf{k}}^{\pi
St,\lambda'\lambda''}(\textbf{p}')&=&(1-\eta_{\pi}(-)^{S+t})\{\delta(p'-k)[\delta(x'-x'')\delta(\varphi'-\varphi'')
\delta_{\lambda'\lambda''}\nonumber
\\&+&\eta_{\pi}(-)^{S}\delta(x'+x'')\delta(\varphi'-\varphi''-\pi)\delta_{\lambda'-\lambda''}]\nonumber
\\&+&\frac{\sum_{N=-S}^{S}e^{iN(\varphi'-\varphi'')}\,d^{S}_{N\lambda'}(x')\,d^{S}_{N\lambda''}(x'')\,}
{2\,d^{S}_{\lambda'\lambda''}(y)}\frac{T^{\pi
St,\lambda'\lambda''}(p',k,y,k^{2})}{k^{2}-p'^{2}+i\varepsilon}\}.
\end{eqnarray}
Eq. (\ref{eq.completness1}) for the singlet and the triplet cases
can be written as:
\begin{eqnarray} \label{eq.completness2}
&&\frac{1}{4}\int_{0}^{\Lambda}dk\,k^{2}\int_{-1}^{1}
d\,x''\,\,\bar{ \Psi}^{\pi 0t,00}_{k,0}}(p',x',x'') {
\,\,\bar{\tilde{\Psi}}_{k,0}^{\pi 0t,00}(p,x'',x)
\nonumber\\&&=(1-\eta_{\pi}(-)^{t})\delta(p'-p)\{\delta(x'-x)
+\eta_{\pi}\delta(x'+x)\},
\end{eqnarray}
\begin{eqnarray} \label{eq.completness3}
&&\frac{1}{4}\int_{0}^{\Lambda}dk\,k^{2}\int_{-1}^{1}
d\,x''\,\,\{2\,\bar{\Psi}^{\pi
1t,\lambda'1}_{k,\lambda'''}}(p',x',x'') {
\,\bar{\tilde{\Psi}}_{k,\lambda'''}^{\pi 1t,1\lambda}(p,x'',x) \nonumber \\
&&+\bar{\Psi}^{\pi 1t,\lambda'0}_{k,\lambda'''}}(p',x',x'') {
\,\bar{\tilde{\Psi}}_{k,\lambda'''}^{\pi 1t,0\lambda}(p,x'',x)\}=
(1+\eta_{\pi}(-)^{t})\delta(p'-p)\nonumber\\&&\times\{\delta(x'-x)
\delta_{\lambda'\lambda}-\eta_{\pi}(-)^{\lambda'''}\delta(x'+x)\delta_{\lambda'-\lambda}\}.
\end{eqnarray}
Finally by considering the vector $\textbf{p}$ along $z$ axis and
using azimuthal behavior of potential and $\omega$-matrix elements,
Eq. (\ref{eq.vlowk}) can be rewritten as:
\begin{eqnarray}
\nonumber V^{\pi St,\lambda'\lambda}_{low\,k}(p',p,x')&=& V^{\pi St,\lambda'\lambda}_{NN}(p',p,x')\\
&+&\frac{1}{4}\sum_{\lambda''}\int_{\Lambda}^{\infty}
dq\,q^{2}\int_{-1}^{1} dx''\,\nu^{\pi
St,\lambda'\lambda''}_{\lambda}(p',q,x',x'')\,\omega^{\pi
St,\lambda''\lambda}(q,p,x''),
\end{eqnarray}
where $x'=\hat{\textbf{p}}'\cdot\hat{\textbf{p}}$ and
$x''=\hat{\textbf{q}}\cdot\hat{\textbf{p}}$. This equation for the
singlet and the triplet cases can be written as:
\begin{eqnarray} \label{eq.vlowk2}
\nonumber V^{\pi 0t,00}_{low\,k}(p',p,x')&=& V^{\pi 0t,00}_{NN}(p',p,x')\\
&+&\frac{1}{4}\int_{\Lambda}^{\infty} dq\,q^{2}\int_{-1}^{1}
dx''\,\,\nu^{\pi 0t,00}_{0}(p',q,x',x'')\,\omega^{\pi
0t,00}(q,p,x''),
\end{eqnarray}
\begin{eqnarray} \label{eq.vlowk3}
\nonumber V^{\pi 1t,\lambda'\lambda}_{low\,k}(p',p,x')&=& V^{\pi 1t,\lambda'\lambda}_{NN}(p',p,x')\\
&+&\frac{1}{4}\int_{\Lambda}^{\infty} dq\,q^{2}\int_{-1}^{1}
dx''\,\,\{2\,\,\nu^{\pi
1t,\lambda'1}_{\lambda}(p',q,x',x'')\,\omega^{\pi
1t,1\lambda}(q,p,x'')\nonumber \\&+&\nu^{\pi
1t,\lambda'0}_{\lambda}(p',q,x',x'')\,\omega^{\pi
1t,0\lambda}(q,p,x'')\}.
\end{eqnarray}

\section{Discussion and Numerical results }\label{sec:Numerical results}
We have chosen AV18 phenomenological potential for our calculations.
This potential is fitted to $pp$ as well as $np$ data below 350
$MeV$ laboratory energy. In addition the AV18 potential is fitted
also to low-energy $nn$ scattering parameters and deuteron
properties. With this interaction in the first step we have
calculated two-body $T$-matrix by solving the LS Eqs.
(\ref{eq.T-matrix1}) and (\ref{eq.T-matrix2}) for the singlet and
the triplet cases respectively \cite{Fachruddin-PRC62}. In the next
step we have calculated $\bar{\tilde{\Psi}}_{k,\lambda'''}^{\pi
St,\lambda''\lambda}(p,x'',x)$ as an inverse of
$\bar{\Psi}_{k,\lambda'''}^{\pi St,\lambda'\lambda''}(p',x',x'')$
for each $\lambda'''$ from Eqs. (\ref{eq.completness2}) and
(\ref{eq.completness3}) by using the LU decomposition method. In the
numerical calculations we have used the Lapack library \cite{60},
for to solve a system of linear equations for the calculation of
$\bar{\tilde{\Psi}}_{k,\lambda'''}^{\pi
St,\lambda''\lambda}(p,x'',x)$ and the two-body $T$-matrix elements.
Then by solving Eqs. (\ref{eq.omega2}) and (\ref{eq.omega3}) we have
obtained $\omega^{\pi St,\lambda'\lambda}(q,p,x)$ and finally we
have inserted the $\omega$-matrix elements into Eqs.
(\ref{eq.vlowk2}) and (\ref{eq.vlowk3}) to obtain the low-momentum
effective interaction $V^{\pi St,\lambda'\lambda}_{low\,k}(p',p,x')$
for the singlet and the triplet cases respectively.

In numerical calculations we have used the Gaussian quadrature grid
points to discrete the momentum and the angle variables. The
integration interval for the $P$ and $Q$ spaces are covered by two
different hyperbolic and linear mappings of the Gauss-Legendre
points from the interval [-1,+1] to the intervals $[0,a]$ and
$[a,b]$:
\begin{eqnarray}
k =\:\frac{1+x}{\frac{1}{\Lambda}-(\frac{1}{\Lambda}-\frac{2}{a})x}
, \qquad\qquad  k = \frac{b-a}{2}x+\frac{b+a}{2}.
\end{eqnarray}
The typical values for $a$ and $b$ are 10 $fm^{-1}$ and 150
$fm^{-1}$ , respectively. As we mentioned in the introduction
section we have used the value of $2.1\,fm^{-1}$ for the cutoff
$\Lambda$ in our calculations. The $\varphi''$-integration within an
interval $[0,2\pi]$ has been rewritten within an interval
$[0,\frac{\pi}{2}]$ as shown in the following notation:
\begin{eqnarray}
I&=&\int^{2\pi}_{0}d\varphi''f(\cos(\varphi'-\varphi''))\,e^{im(\varphi'-\varphi'')}=
\int^{2\pi}_{0}d\varphi''f(\cos\varphi'')\,e^{im\varphi''}\nonumber\\
&=&\int^{\pi}_{0}d\varphi''\{f(\cos\varphi'')\,e^{im\varphi''}+f(-\cos\varphi'')\,e^{im(\varphi''+\pi)}\}\nonumber\\
&=&\int^{\,\frac{\pi}{2}}_{0}d\varphi''\{f(\cos\varphi'')(e^{im\varphi''}+e^{im(2\pi-\varphi'')})+
f(-\cos\varphi'')(e^{im(\pi+\varphi'')}+e^{im(\pi-\varphi'')}\}.
\end{eqnarray}
The second equality has been justified by the periodicity of the
integrand within $2\pi$. Thus the number of integration points for
polar angle has been reduced.

In our calculations we have chosen sixty grid points for the
momentum variables in the interval $[0,a]$, and twenty two grid
points for the momentum variables in the interval $[a,b]$. Also
forty and ten grid points for the spherical and the polar angle
variables have been used respectively. The solutions of the integral
Eqs. (\ref{eq.T-matrix1}) and (\ref{eq.T-matrix2}) require a
one-dimensional interpolation. We have used the cubic hermitian
splines of Ref. \cite{59} for its accuracy and high computational
speed.

\begin{figure}
\includegraphics[width=12cm]{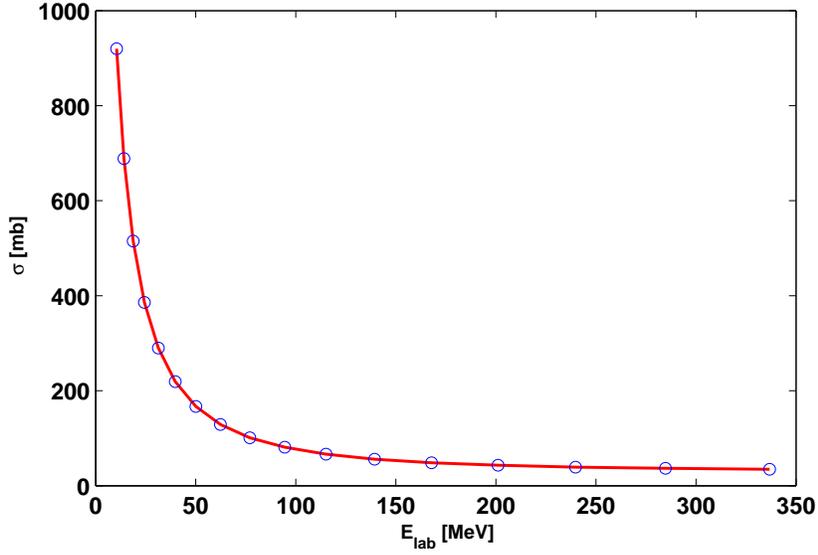}
\caption{\label{fig.cross section}The total cross section from the
low-momentum effective potential $V_{low\,k}$ (circles) and the bare
potential $V_{NN}$ (solid line) as a function of kinetic energy in
the lab frame.}
\end{figure}

\begin{figure}
\includegraphics[width=15cm]{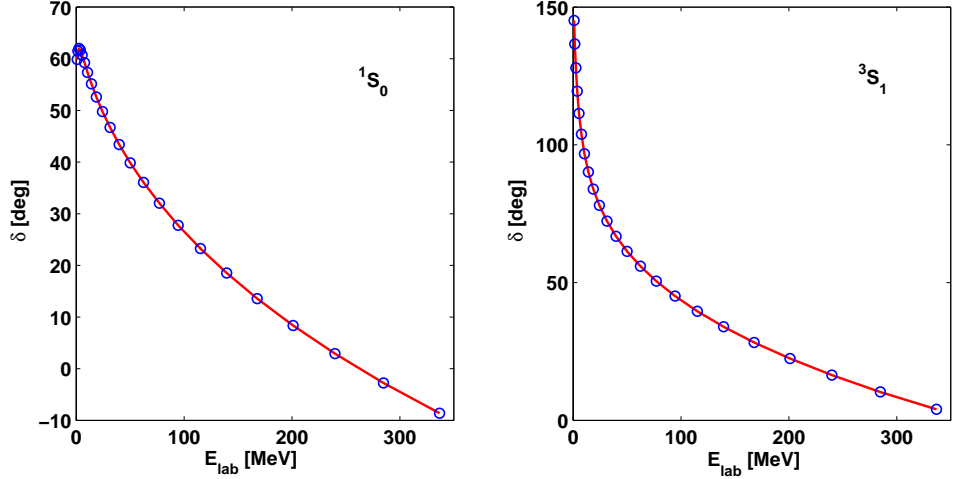}
\caption{\label{fig.phase shifts1}The $\,^{1}S_{0}$ and
$\,^{3}S_{1}$-wave phase shifts from the low-momentum effective
potential $V_{low\,k}$ (circles) and the bare potential $V_{NN}$
(solid line) as a function of kinetic energy in the lab frame.}
\end{figure}

In Fig. \ref{fig.cross section} we have shown the total two-body
cross section for $np$ scattering by using $V_{low\,k}$ and
$V_{NN}$. Also in Fig. \ref{fig.phase shifts1} we have compared the
calculated $\,^{1}S_{0}$ and $\,^{3}S_{1}$-wave phase shifts from
$V_{low\,k}$ and $V_{NN}$. For calculation of the phase shifts we
have used the relation between the PW and 3D representations of the
on-shell two-body $T$-matrix \cite{Fachruddin-PRC68}:
\begin{eqnarray}
T^{Sjt,l'l}(p,p,p^{2})&=&\frac{\pi}{2}\frac{\sqrt{2l'+1}\sqrt{2l+1}}{2j+1}
\sum_{\lambda'\lambda}C(l'Sj;0\lambda')\,C(lSj;0\lambda)\nonumber
\\&\times&\int_{-1}^{1}dx'\,\,\,d^{j}_{\lambda\lambda'}(x')\,\,
T^{\pi St,\lambda'\lambda}(p,p,x',p^{2}).
\end{eqnarray}

\begin{figure}
\includegraphics[width=14cm]{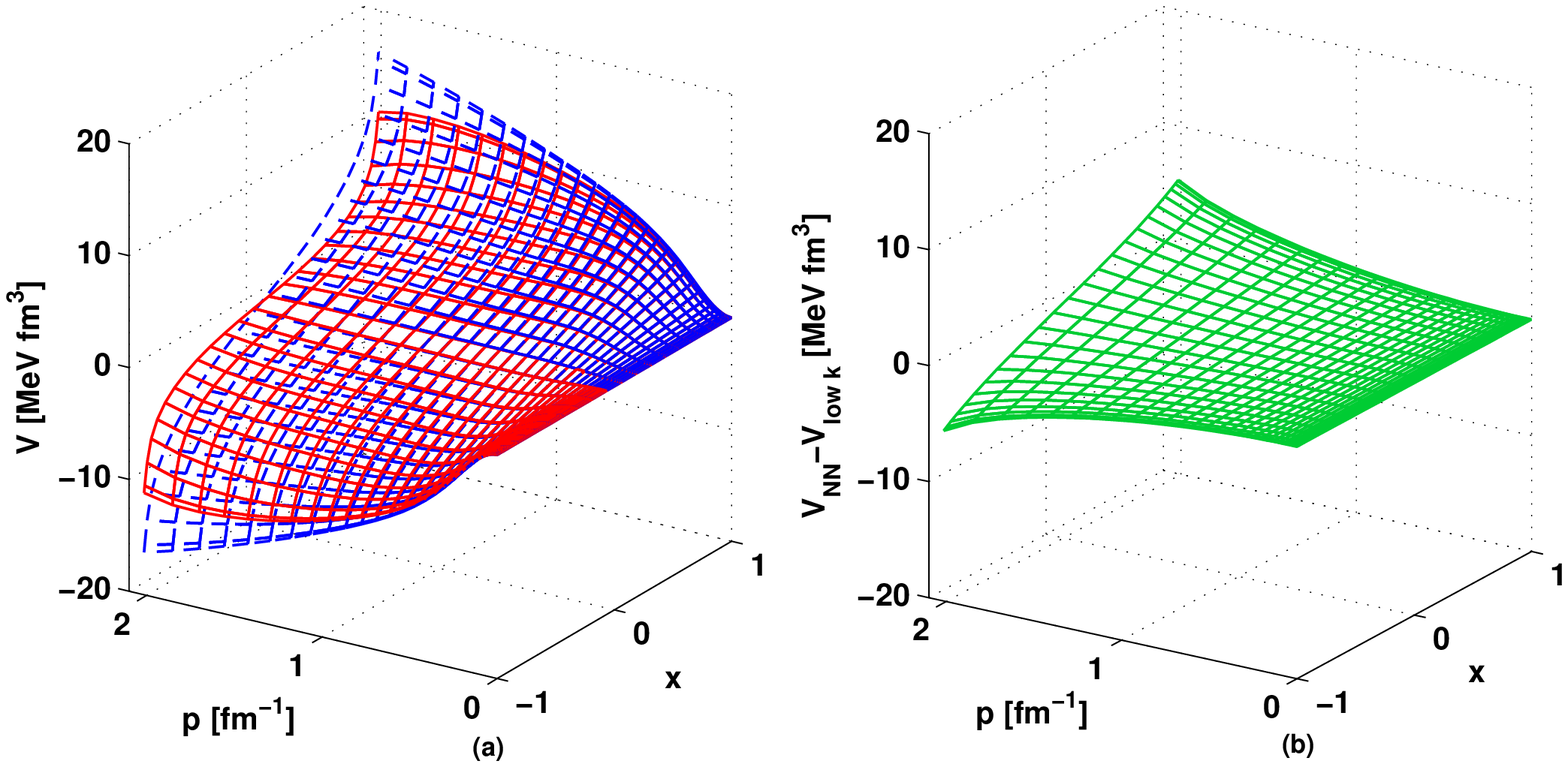}
\caption{\label{fig.vlowkS0t0} (a) The comparison of the
low-momentum effective potential $V_{low\,k}$ (solid lines) with the
bare potential $V_{NN}$ (dashed lines) and (b) differences between
them, for $S=0$ and $t=0$.}
\end{figure}

\begin{figure}
\includegraphics[width=14cm]{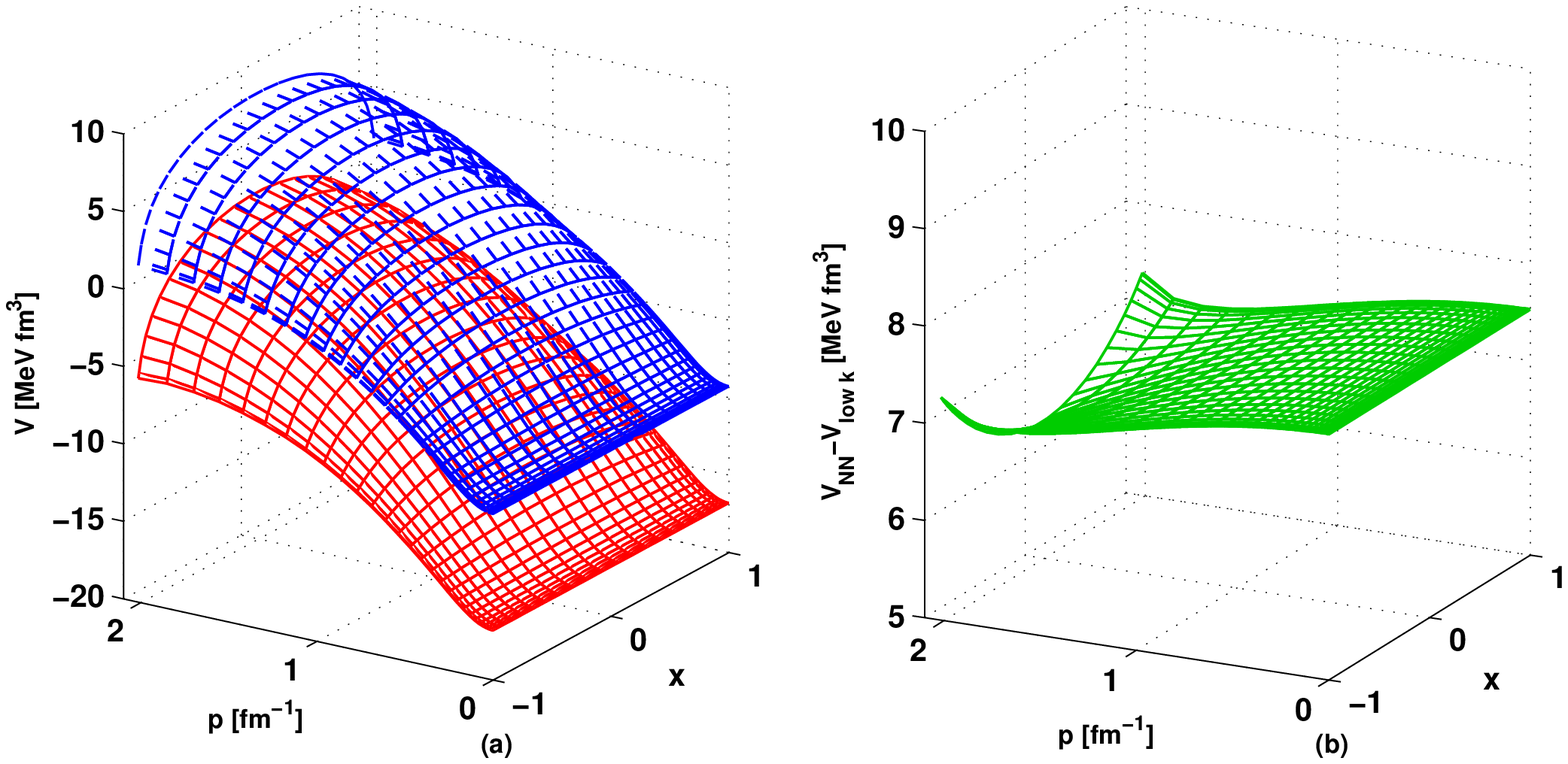}
\caption{\label{fig.vlowkS0t1} (a) The comparison of the
low-momentum effective potential $V_{low\,k}$ (solid lines) with the
bare potential $V_{NN}$ (dashed lines) and (b) differences between
them, for $S=0$ and $t=1$.}
\end{figure}

\begin{figure}
\includegraphics[width=15cm]{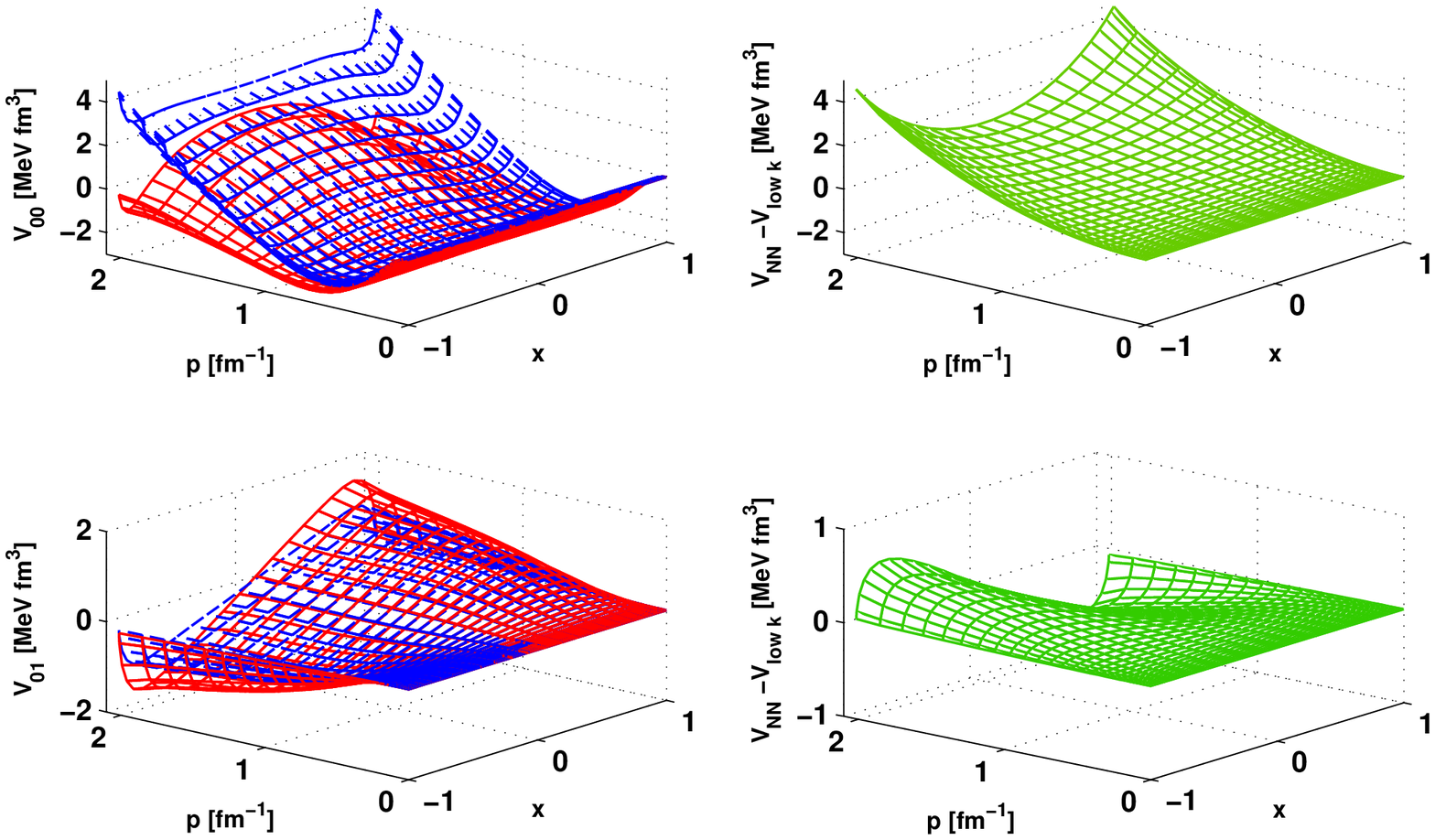}
\\
\includegraphics[width=15cm]{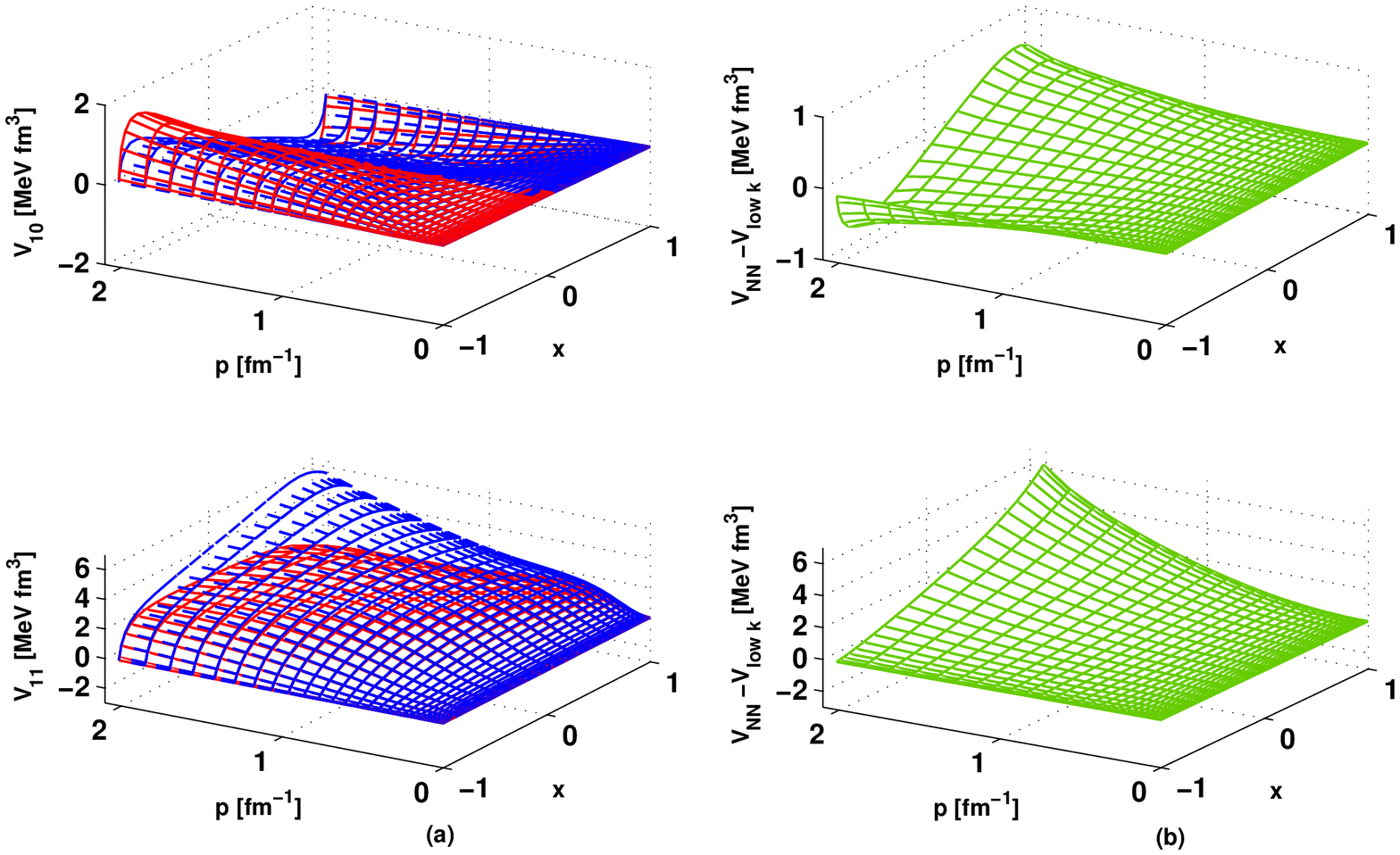}
\caption{\label{fig.vlowkS1t0} (a) The comparison of the
low-momentum effective potential $V_{low\,k}$ (solid lines) with the
bare potential $V_{NN}$ (dashed lines) and (b) differences between
them, for $S=1$ and $t=1$.}
\end{figure}

\begin{figure}
\includegraphics[width=15cm]{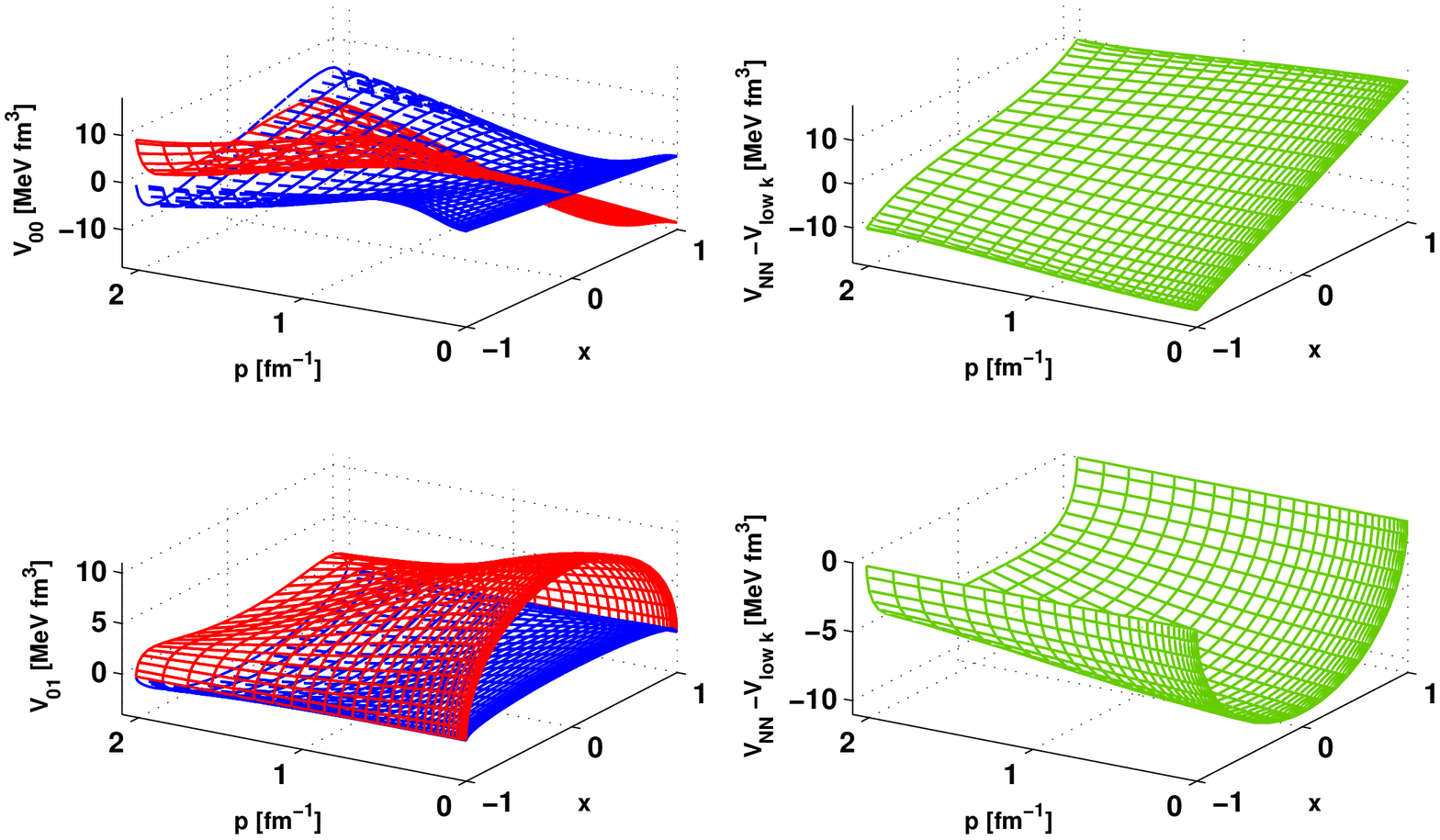}
\\
\includegraphics[width=15cm]{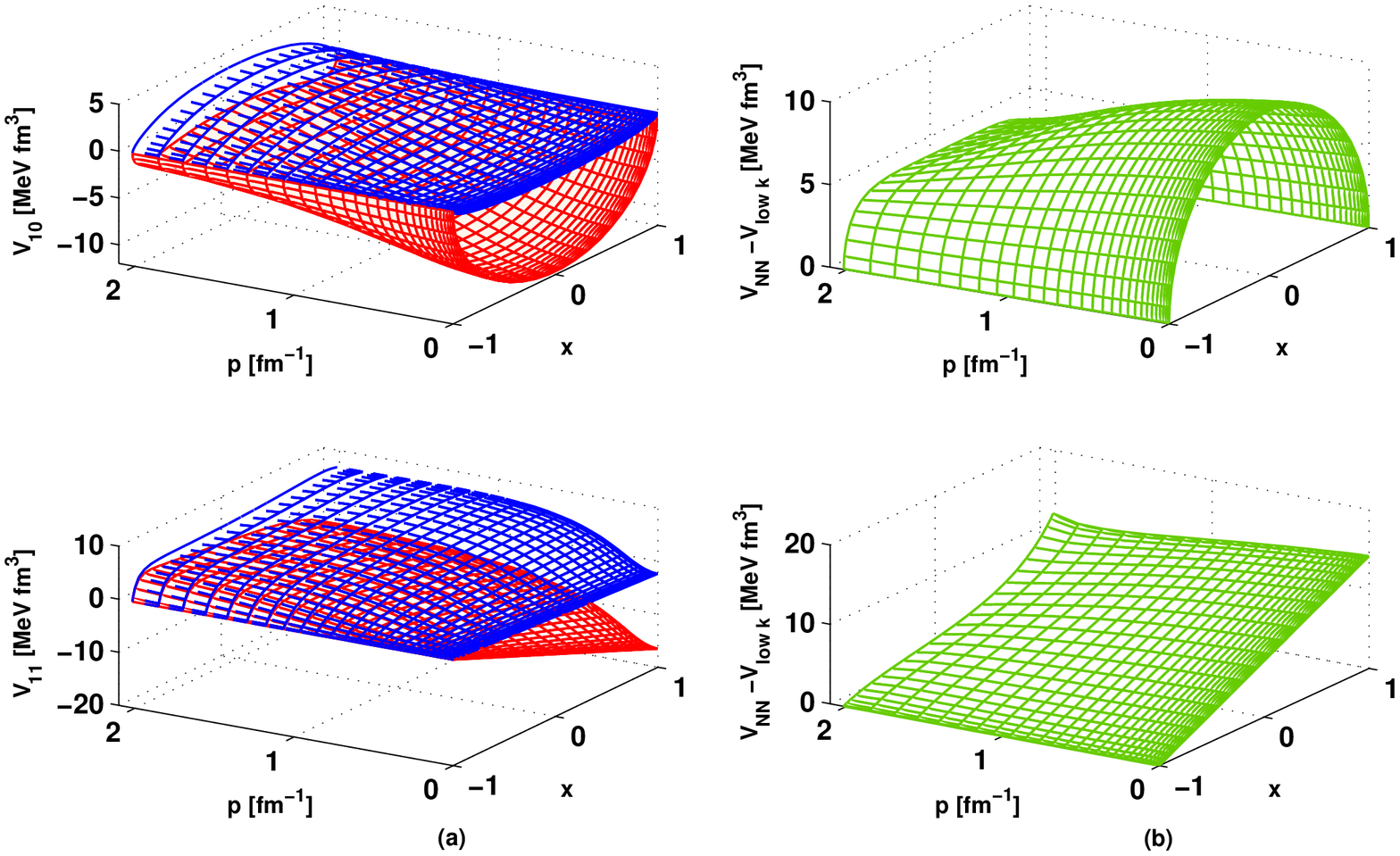}
\caption{\label{fig.vlowkS1t1} (a) The comparison of the
low-momentum effective potential $V_{low\,k}$ (solid lines) with the
bare potential $V_{NN}$ (dashed lines) and (b) differences between
them, for $S=1$ and $t=0$.}
\end{figure}

The results are in good agreement with high accuracy. In Figs.
{\ref{fig.vlowkS0t0}-{\ref{fig.vlowkS1t1} the calculated
low-momentum effective potential $V^{\pi
St,\lambda'\lambda}_{low\,k}(p,p,x')$ and the bare potential $V^{\pi
St,\lambda'\lambda}_{NN}(p,p,x')$ as well as differences between
them have been shown as a function of the momentum variable $p$ and
the angle variable $x$.

As a test of our calculations we have compared the obtained results
for the low-momentum effective potential and the $\omega$ operator
in the 3D and the PW approaches. As a first step we have calculated
the $\omega$ operator and the low-momentum effective potential in
the PW approach directly. We have then obtained the PW projection of
the $\omega$ operator and the low-momentum effective potential from
their corresponding 3D representation by the following relations:

\begin{eqnarray}
V_{low\,k}^{Sjt,l'l}(p,p)&=&\frac{\pi}{2}\frac{\sqrt{2l'+1}\sqrt{2l+1}}{2j+1}
\sum_{\lambda'\lambda}C(l'Sj;0\lambda')\,C(lSj;0\lambda)\nonumber
\\&\times&\int_{-1}^{1}dx'\,\,\,d^{j}_{\lambda\lambda'}(x')\,\,
V_{low\,k}^{\pi St,\lambda'\lambda}(p,p,x'),  \\
\omega^{Sjt,l'l}(q,p)&=&\frac{\pi}{2}\frac{\sqrt{2l'+1}\sqrt{2l+1}}{2j+1}
\sum_{\lambda'\lambda}C(l'Sj;0\lambda')\,C(lSj;0\lambda)\nonumber
\\&\times&\int_{-1}^{1}dx\,\,\,d^{j}_{\lambda\lambda'}(x)\,\,
\omega^{\pi St,\lambda'\lambda}(q,p,x).
\end{eqnarray}

\begin{figure}
\includegraphics[width=14cm]{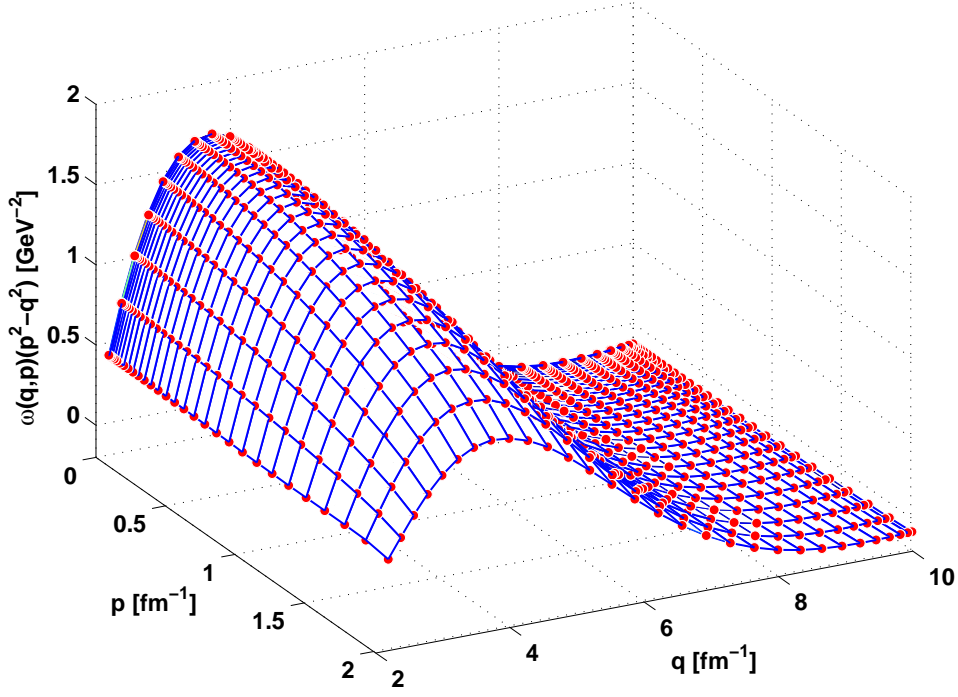}
\caption{\label{fig.Omega-Swave}The comparison of
$\omega(q,p)(p^{2}-q^{2})$ for the $^{1}S_{0}$-wave calculated in
the PW approach (dotes) and in the 3D approach (solid lines).}
\end{figure}

\begin{figure}
\includegraphics[width=14cm]{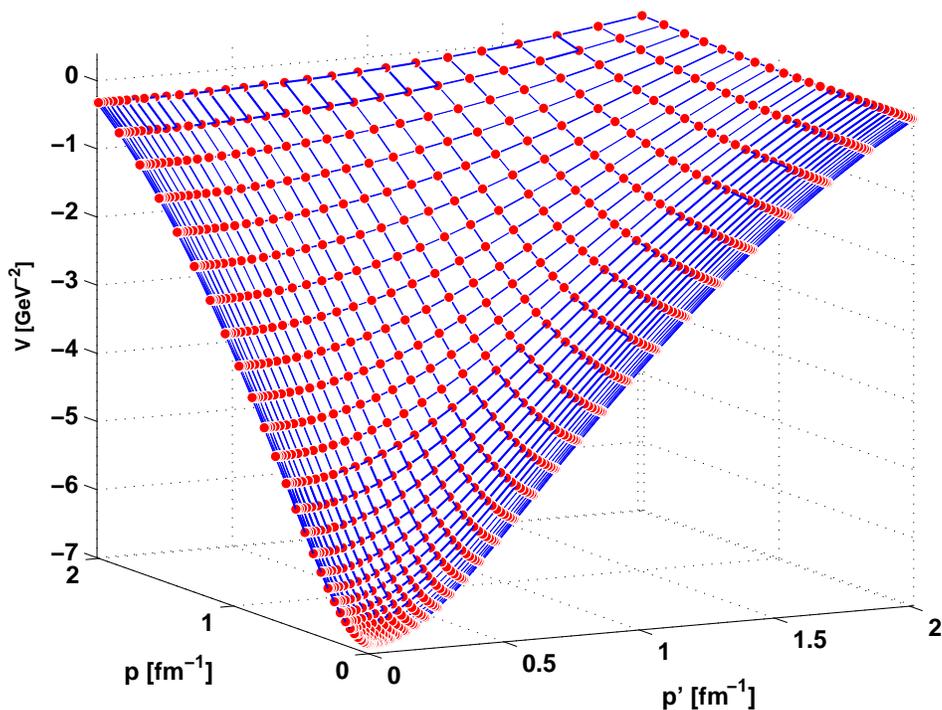}
\caption{\label{fig.Vlowk-Swave}The comparison of the low-momentum
effective potential $V_{_{low\,k}}$ for the $\,^{1}S_{0}$-wave
calculated in the PW approach (dotes) and in the 3D approach (solid
lines).}
\end{figure}

The obtained results for the channel$\,^{1}S_{0}$ in the 3D and PW
approaches have been given in Fig. {\ref{fig.Omega-Swave} and Fig.
{\ref{fig.Vlowk-Swave}. The agreement between the two approaches is
quite satisfactory.

\section{Summary and Outlook}\label{sec:summary}
In this article the 3D formulation of the model space Lee-Suzuki
method based on momentum-helicity representation has been presented
on momentum-helicity basis states. The low-momentum effective
interaction $V_{low\,k}$ has been derived as a function of the
magnitude of momenta and the angle between them for the singlet and
the triplet cases respectively without using the partial wave
decomposition. The calculated two-body observables from the
low-momentum effective interaction and the bare interaction have
been presented. In addition, a comparison between the calculated
$V_{low\,k}$ from the PW and the 3D approach has been demonstrated
as a test of our calculations.

The advantage of our formulation in the 3D representation in
comparison with the PW representation is that we have calculated the
low-momentum effective interaction by considering all partial waves
automatically. The implementation of the obtained 3D low-momentum
effective interaction in the few-body bound and scattering
calculations is a major task that can be considered for the future
investigations.

\section*{Acknowledgments}
This work was supported by the research council of the University of
Tehran.

\end{document}